\documentclass[twocolumn,preprintnumbers,amsmath,amssymb,showpacs]{revtex4}

\usepackage{graphicx}% Include figure files
\usepackage{dcolumn}% Align table columns on decimal point
\usepackage{bm}% bold math

\begin{document}

\title{Excess Vibrational Modes and the Boson Peak in Model Glasses}

\author{Ning Xu$^{1,2}$}
\author{Matthieu Wyart$^3$}
\author{Andrea J. Liu$^1$}
\author{Sidney R. Nagel$^2$}

\affiliation{$^1$~Department of Physics and Astronomy, University of
Pennsylvania, Philadelphia, PA 19104.\\
$^2$~James Franck Institute, The University of Chicago, Chicago, IL 60637.\\
$^3$~Department of Physics, Harvard University, Cambridge, MA 02138.}

\date{\today}

\begin{abstract}

The excess low-frequency normal modes for two widely-used models of
glasses were studied at zero temperature.  The onset frequencies for
the anomalous modes for both systems agree well with predictions of a
variational argument, which is based on analyzing the vibrational
energy originating from the excess contacts per particle over the
minimum number needed for mechanical stability.  Even though both
glasses studied have a high coordination number, most of the
additional contacts can be considered to be weak.
\end{abstract}


\maketitle

Our understanding of liquids is based on the idea that liquid
structure is largely determined by strong short-ranged repulsions and
that the longer-ranged attractions can be treated as a
perturbation~\cite{weeks}.  Similar considerations were used to study
jamming at zero temperature as a function of density
$\phi$~\cite{epitome,silbert}.  When only finite-ranged repulsions are
included, so that non-overlapping particles do not interact, there is
a sharp jamming transition at $\phi_c$.  In such systems, the
low-frequency normal modes of vibration in the marginally jammed state
are fundamentally different from the long-wavelength plane waves
expected from elasticity theory~\cite{epitome,silbert}.  The unusual
nature of the modes is reflected in the density of vibrational states,
$D(\omega)$, versus frequency, $\omega$~\cite{epitome,silbert}.  At
the transition, $D(\omega)$ has a plateau extending to $\omega=0$.
This is very different from the expected Debye scaling $D(\omega)
\propto \omega^{d-1}$ in $d$ dimensions~\cite{ashcroftmermin}, which
is normally observed in solids.  At densities above $\phi_c$, the
plateau no longer extends to $\omega=0$ but terminates at a frequency
$\omega^*$.  The dramatic rise in $D(\omega)$ at $\omega^*$
corresponds to the onset of anomalous modes.  Similar excess modes,
whose origins are still
debated~\cite{fractal,duval,keyes,schirmacher,kan,grigera1,grigera2,gurevich},
are observed in glasses.  The anomalous modes in glasses give rise to
what is known as the boson
peak~\cite{sokolov,wuttke,lunkenheimer,nakayama}, and are believed to
be responsible for many characteristic low-temperature
phenomena~\cite{phillips}.

In this paper, we demonstrate that the theoretical
framework~\cite{wyartepl,wyartpre,wyartthesis} used to explain the
modes in marginally jammed solids at zero temperature can be extended
to the anomalous modes in two more realistic models of ``repulsive
glasses," by which we mean systems that undergo glass transitions due
to the repulsive part of their potentials as the temperature is
lowered.  The zero-temperature jamming transition of spheres with
finite-ranged repulsions, which we will call Point J to distinguish
from other jamming transitions, coincides with the density at which
the system has just enough contacts to satisfy the constraints of
mechanical equilibrium~\cite{epitome}.  This is called the isostatic
condition.  The average coordination number $z$ ({\it i.e.} the
average number of particles with which a given particle interacts)
needed for mechanical stability is $z_c = 2d$, where $d$ is the
spatial dimension ~\cite{maxwell,shlomo}.  By compressing the system
or increasing the range of interaction, we increase $z$.  These extra
contacts suppress anomalous modes at low
frequencies~\cite{wyartepl,wyartpre,wyartthesis}.  We will show that
in systems with long-ranged interactions, there is a well-defined
division between a relatively few strong repulsive interactions (stiff
contacts) and the more numerous weaker interactions (including
attractions), which can be treated as a correction.

We will study the onset frequency of the anomalous modes, $\omega^\dagger$,
in two widely-used models of glasses and the glass transition and
compare the simulation results with theoretical
predictions~\cite{wyartthesis}.  We use a mixture of $800$ A and $200$
B spheres with equal mass $m$ interacting in three dimensions via the
Lennard-Jones potential \cite{kob}:
\begin{equation}
V(r_{ij})= \frac{\epsilon_{ij}}{72} \left[
\left(\frac{\sigma_{ij}}{r_{ij}}\right)^{12} -
\left(\frac{\sigma_{ij}}{r_{ij}}\right)^6\right], \label{lj}
\end{equation}
where $r_{ij}$ is the separation between particles $i$ and $j$ and
$\epsilon_{ij}$ and $\sigma_{ij}$ depend on the type of particles
under consideration: $\epsilon_{AB} = 1.5\epsilon_{AA}$,
$\epsilon_{BB} = 0.5\epsilon_{AA}$ and $\sigma_{AB} = 0.8
\sigma_{AA}$, and $\sigma_{BB} = 0.88\sigma_{AA}$.  The potential is
cut off at $r_{ij}=2.5\sigma_{ij}$ and shifted to satisfy
$V(2.5\sigma_{ij})=V^{\prime}(2.5\sigma_{ij})=0$. For these systems,
the density is given in units of $\rho=N\sigma_{AA}^3/L^3$ and
$\epsilon_{AA}$, $\sigma_{AA}$ and $m$ are set to unity.  We restrict
the densities to lie above $\rho \approx 1.2$, where the pressure is
positive.  When this is not the case, there can be low-frequency modes
arising from rather different physics~\cite{shlomo}.

For comparison we also study a system with purely repulsive
interactions where $\phi_c$ exists.  We simulate a mixture of $500$ A
and $500$ B spheres with $\sigma_B=1.4\sigma_A$ and equal mass $m$,
interacting via the repulsive Lennard-Jones potential \cite{weeks}
\begin{equation}
V(r_{ij}) =
\begin{cases}
\frac{\epsilon}{72} \left[ \left(
\frac{\sigma_{ij}}{r_{ij}}\right)^{12} - 2 \left(
\frac{\sigma_{ij}}{r_{ij}}\right)^6 + 1\right], &
\text{$r_{ij}<\sigma_{ij}$,} \\ 0, & \text{$r_{ij}\ge \sigma_{ij}$,}
\end{cases}\label{rlj}
\end{equation}
where $\epsilon$ is the characteristic energy, and
$\sigma_{ij}=(\sigma_i+\sigma_j)/2$.  For these systems, we
characterize density with the packing fraction $\phi=\pi/6 (N_A
\sigma_A^3 + N_B \sigma_B^3)/L^3$.  We set $\epsilon=1$, $m=1$, and
$\sigma_A=1$.  The simulations were all carried out in a cubic box
with periodic boundary conditions.  We study zero-temperature ($T=0$)
configurations which were obtained by quenching initially random
($T=\infty$) configurations of particles to their local energy minima
at $T=0$ using conjugate gradient energy minimization \cite{fortran}.

In the harmonic expansion, the energy of particle displacements
$\delta \vec R_i$ from their equilibrium positions is:
\begin{equation}
\delta E=\frac{1}{2} \sum_{i,j}\biggl [V^{\prime\prime}(r_{ij})
(\delta \vec R_{ij}\cdot \hat r_{ij})^2 +
\frac{V^{\prime}(r_{ij})}{r_{ij}} (\delta \vec R_{ij}^\perp)^2 \biggr
]
\label{dEexpn}
\end{equation}
where $V^{\prime}$ and $V^{\prime\prime}$ are the first and second
derivatives of the pair potential $V(r)$ with respect to $r$, and
$\vec r_{ij}= r_{ij} \hat r_{ij}$ is the equilibrium separation vector
between particles $i$ and $j$.  Here, $\delta \vec R_{ij}\equiv \delta
\vec R_i-\delta \vec R_j$, and $\delta \vec R_{ij}^\perp$ is the
projection of $\delta \vec R_{ij}$ on the plane perpendicular to $\hat
r_{ij}$.  We diagonalize the dynamical matrix~\cite{ashcroftmermin} to
obtain normal modes $|n \rangle$, and their corresponding eigenvalues
${\cal E}_n$.  The normal-mode frequencies are $\omega_n^2 = {\cal
E}_n$, where the index $n$ runs from $1$ to $Nd$.  From these
frequencies,we calculate the density of vibrational states per
particle, $D(\omega)$.  It is instructive to also calculate
$D^{(0)}(\omega)$, obtained by neglecting the second term in
Eq.~\ref{dEexpn}({\it i.~e.} the stress term).  This corresponds to
replacing $V(r)$ with unstretched springs, each chosen to have the
same stiffness $V^{\prime \prime}(r)$ as for the original potential.

\begin{figure}
\scalebox{0.5}{\includegraphics{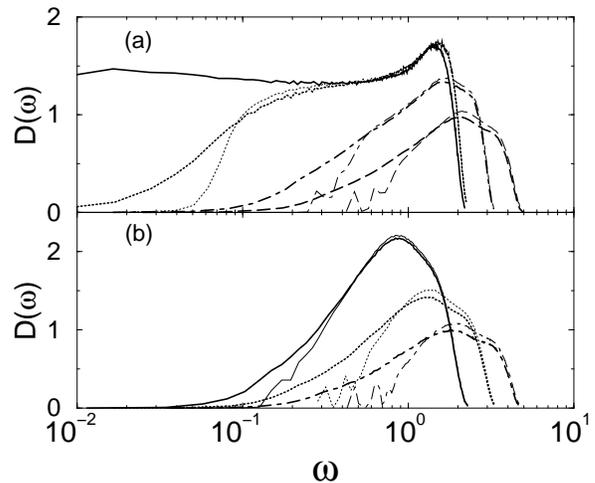}}%
\caption{\label{fig:RLJ} Density of vibrational states per particle of
3D glasses of $N=1000$ particles.  We plot $D(\omega)$ for stressed
systems (heavy curves) and $D^{(0)}(\omega)$ for unstressed systems
(light curves) interacting via (a) the repulsive Lennard-Jones
potential at $\Delta\phi=10^{-6}$ (solid), $10^{-2}$ (dotted), $0.1$
(dot-dashed), and $0.2$ (dashed); and (b) the Lennard-Jones potential
at $\rho=1.2$ (solid), $1.4$ (dotted) and $1.6$ (dashed).}
\end{figure}

In Fig.~\ref{fig:RLJ}(a) and (b), we show two curves, $D(\omega)$
(heavy curves) and $D^{(0)}(\omega)$ (light curves), corresponding to
the stressed and unstressed cases, at each density.
Fig.~\ref{fig:RLJ}(a) shows results for the repulsive Lennard-Jones
systems.  Close to the unjamming transition, these systems are
approximately equivalent to the repulsive harmonic systems studied
previously~\cite{epitome,silbert}.  Just above the transition,
$D(\omega)$ has a plateau down to $\omega=0$.  Upon compression the
onset of these anomalous modes shifts up to $\omega^*$, below which
$D(\omega)$ decreases to zero.  Fig. \ref{fig:RLJ}(b) shows the
results for the system with Lennard-Jones interactions.  Because there
are attractive interactions, the jamming transition lies inside the
liquid-vapor spinodal~\cite{sastry,epitome} and is inaccessible.
Thus, the plateau in the density of states never extends to $\omega =
0$.

%%%%%%%%%%%%%%%%%
%%%%%%%%%%%%%%%%%

It is important to note that $D^{(0)}(\omega)$ falls much more sharply
than $D(\omega)$ as $\omega$ decreases below $\omega^*$, but that
$D(\omega)$ and $D^{(0)}(\omega)$ are nearly indistinguishable above
$\omega^*$.  Here, $\omega^*$ corresponds to the onset of the
anomalous modes in unstressed systems.  At high densities, the sharp
peaks in $D^{(0)}(\omega)$ below $\omega^*$ arise from linear
combinations of nearly-degenerate long-wavelength plane waves.  (The
lowest peak has plane waves with wavevector, $k_1=2\pi/L$ where $L$ is
the box size, while the second peak has $k_2=2\sqrt2\pi/L$ as one
would expect for the two lowest frequency modes in an elastic solid.)
Thus, the onset of anomalous modes lies above these peaks.  In the
infinite system-size limit, the peaks should smooth out to yield the
normal scaling: $D(\omega) \sim \omega^{d-1}$ in $d$ dimensions.

We now recapitulate the theoretical ideas that address the properties
of high-coordination systems~\cite{wyartthesis}.  In general, extra
contacts increase the frequency of the lowest-frequency anomalous
modes in two ways: (i) they can increase the energy cost of a mode by
adding extra nodes so that some bonds are unstretched during an
oscillation, and (ii) they can leave the number of nodes fixed but
instead increase the average restoring force (and therefore the
energy) for the normal-mode displacement.

A variational argument calculates an upper bound for the energy of a
normal mode by minimizing the energy with respect to these two
contributions.  The first term in Eq.~\ref{dEexpn} indicates that a
good trial function would have nodes, {\it i.~e.,} small values of
$(\delta \vec R_{ij} \cdot \hat r_{ij})^2$, where $V^{\prime\prime}$
is large.  Thus the $z_1 N/2$ contacts with the highest values of
$V^{\prime\prime}$ introduce nodes in the trial function.  The
remaining $(z-z_1) N/2$ contacts increase the energy of the trial mode
by increasing the restoring force for displacements according to the
first term in Eq.~\ref{dEexpn}~\cite{wyartthesis}.  We rewrite
Eq.~\ref{dEexpn} as
\begin{eqnarray}
\delta E^{(0)} &=& \delta E_1+\delta E_2, \\ \nonumber \delta E &=&
\delta E^{(0)}+ \delta E_3, \label{dE123}
\end{eqnarray}
where $\delta E_1$, $\delta E_2$ and $\delta E_3$ represent the energy
costs associated respectively with the $z_1N/2$ stiffest contacts, the
remaining $(z-z_1)N/2$ weak contacts, and the contribution of the
stress term (the second term in Eq.~\ref{dEexpn}; see
Ref.~\cite{wyartpre}).  Note that $\delta E^{(0)}$ is the energy cost
of a mode for the unstressed system.

We now construct approximate expressions for these contributions.
Ref.~\cite{wyartepl} argues that $\delta E_1 \approx A_1 k_1
(z_1-z_c)^2$, where $k_1= \langle V^{\prime\prime} \rangle_1$ is the
average over the $z_1 N/2$ contacts with the highest values of
$V^{\prime\prime}$.  This is consistent with earlier simulation
results for harmonic springs for $z_1-z_c \leq 3$~\cite{wyartpre}.
While $k_1$ varies strongly with density and potential, we expect
$A_1$ to depend only weakly on these quantities~\cite{A1dep}.  To
obtain $A_1$, we compare to simulations of unstressed systems with
springs of equal stiffness.  The precise value of $A_1$ depends on
which point in the density of states we choose to represent the onset
of anomalous modes.  In the following analysis, we use the value
$\omega^\dagger$ where $D^{(0)}(\omega)$ reaches 0.25 of its maximum
height, which fixes $A_1=0.018$.  We would have obtained a somewhat
different constant if we had compared with data at a different point
in the rise, but the difference would be small because
$D^{(0)}(\omega)$ rises abruptly.

To estimate $\delta E_2$, we assume that for the normalized trial
mode, the displacements of $i$ and $j$ are uncorrelated with each
other if $i$ and $j$ are connected by weak contacts. Then $\langle
(\delta \vec R_{ij} \cdot \hat r_{ij})^2 \rangle \approx 2/Nd$ for an
$N$-particle system in $d$ dimensions~\cite{wyartthesis}.  Thus
$\delta E_2 \approx \frac{1}{Nd} \sum_{ij}^w
V^{\prime\prime}(r_{ij})$, where the sum $\sum^w$ runs only over pairs
of particles $i$ and $j$ connected by the $(z-z_1)N/2$ weakest
contacts. We have shown numerically that this approximation is
reasonable for a system with particles connected by harmonic springs
at their equilibrium lengths with two very different stiffnesses.  For
such a system, the stiff springs contribute only to $\delta E_1$, the
weak springs contribute only to $\delta E_2$.  Our expressions for
$\delta E_2$ and $\delta E_1$ were thus verified cleanly.

We estimate $\delta E_3$ as follows. For uncorrelated displacements
between particles $i$ and $j$, $\langle (\delta \vec R_{ij}^\perp)^2
\rangle \approx 2(d-1)/dN$, leading to $\delta E_3 = A_3 \cal{S}$,
where ${\cal S} \equiv 1/N \sum_{ij}
\frac{V^{\prime}(r_{ij})}{r_{ij}}$, and $A_3 \approx (d-1)/d$ is of
order unity.

Note that $\delta E_3$ does not depend on $z_1$, so it does not affect
the energy minimization with respect to $z_1$ or $z_1$ itself. In
order to compute $z_1$, we thus compute the total energy cost in
absence of stress:
\begin{equation}
\delta E^{(0)}=0.018 k_1 (z_1-z_c)^2+\frac{1}{Nd} \sum_{ij}^w V^{\prime\prime}(r_{ij}),
\label{dEfinal}
\end{equation}
where $k_1= \langle V^{\prime\prime} \rangle_1$ is the average over
the $z_1 N/2$ contacts with the highest values of $V^{\prime\prime}$,
and $\sum_{ij}^w$ sums over the pairs of particles connected by the
$(z-z_1)N/2$ contacts with the smallest values of $V^{\prime\prime}$.
By minimizing $\delta E^{(0)}$ with respect to $z_1$, we obtain an
upper bound on the energy and therefore the frequency,
$\omega^*=\sqrt{\delta E_{\rm min}}$, of the lowest-frequency
anomalous modes in the unstressed systems.
   
%%%%%%%%%%%%%%%%%
\begin{figure}
\scalebox{0.5}{\includegraphics{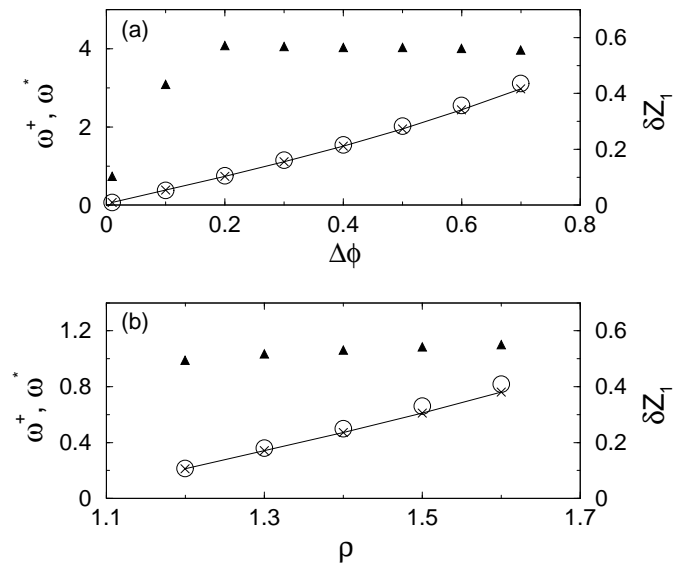}}%
\caption{\label{fig:omega} {\bf Left axis}: The characteristic
frequency $\omega^\dagger$ (open symbols) calculated from numerical
simulations and the corresponding theoretical predictions $\omega^*$
(crosses).  Lines are to guide the eye and connect the theoretical
points.  {\bf Right axis}: the fractional deviation from isostaticity
$\delta Z_1 \equiv (z_1-z_c)/z_c$ (closed symbols), as functions of
(a) distance from the unjamming transition, $\phi-\phi_c$, for
repulsive Lennard Jones mixtures and (b) density $\rho$ for
Lennard-Jones mixtures.}
\end{figure}
%%%%%%%%%%%%%%%%%

In Fig.~\ref{fig:omega}, we compare this theoretical prediction for
$\omega^*$ with the onset of anomalous modes, $\omega^\dagger$, from
simulations in the unstressed system (determined by where
$D^{(0)}(\omega)$ reaches 0.25 of its maximum value).  The agreement
is excellent everywhere, with no adjustable parameters.

Our results show that even though a typical repulsive amorphous solid
may have a high coordination number and therefore appear to be far
from the unjamming transition, most of the contacts are weak.  The
number of {\it stiff} contacts is only slightly in excess of the
minimum needed for mechanical stability.  This is shown in
Fig.~\ref{fig:omega} (solid triangles) for both potentials at all
densities studied. Even for the Lennard-Jones system, where the
power-law tail of the interaction leads to a divergent total
coordination, we find that $(z_1-z_c)/z_c <0.6$ at all densities. Thus
$(z_1-z_c)/z_c$ is a small parameter.

\begin{figure}
\scalebox{0.5}{\includegraphics{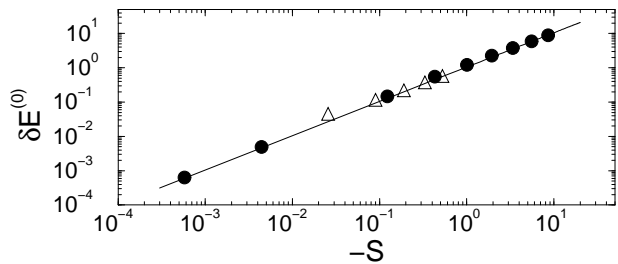}}%
\caption{\label{fig:dEplot} The energy cost of the lowest frequency
anomalous mode for an unstressed system, $\delta E^{(0)}$, as a
function of ${\cal S} \equiv 1/N \sum_{ij}
{V^{\prime}(r_{ij})/r_{ij}}$ for repulsive Lennard-Jones mixtures
(solid circles) and Lennard-Jones mixtures (open triangles).  The
contribution of the stress term to the energy cost of a mode is
$\delta E_3=A_3 {\cal S}$, where $A_3$ is of order unity.  The
straight line fit corresponds to $\delta E^{(0)}= -1.6 {\cal S}$.}
\end{figure}

While the results of Fig.~\ref{fig:omega} are for unstressed systems,
real glasses have nonzero stresses.  To estimate the onset of
anomalous modes for systems with stress, we must calculate the total
energy cost of a mode, $\delta E=\delta E^{(0)}+\delta E_3$.
Fig.~\ref{fig:dEplot} shows that for our systems, $\delta E^{(0)}
\approx -\delta E_3$, where $\delta E^{(0)}$ is evaluated at its
minimum with respect to $z_1$ for both potentials at different
densities.  Thus, there is a near cancellation of two large terms
leading to a small value of $\delta E$ (with a large uncertainty) and
therefore a very low onset frequency for the anomalous modes in the
stressed systems.  This is consistent with the finding that the
expected scaling of $D(\omega) \sim \omega^2$ for plane-wave normal
modes is eclipsed by an approximately linear frequency dependence at
small $\omega$ for both potentials.  The near-cancellation may be a
result of the history of how the system was prepared~\cite{wyartpre}.

%%%%%%%%%%%%%%%%%
%%%%%%%%%%%%%%%%%

Our results provide a plausible explanation for the origin of the
excess vibrational modes of the boson peak in repulsive glasses.  For
two commonly-studied models, we have shown that the boson peak derives
from the same low-frequency anomalous modes that arise at Point J for
systems with finite-ranged repulsions.  Several theories have been
advanced previously for the boson
peak~\cite{fractal,duval,keyes,schirmacher,kan,grigera1,grigera2,gurevich}.
Fractal systems~\cite{fractal} as well as disordered
ones~\cite{schirmacher,kan,grigera1} can exhibit excess vibrational
modes, but glasses are typically not fractal and low-coordination
crystals can also display excess modes~\cite{nakayama}.  Approaches by
Phillips~\cite{jcphillips}, Thorpe~\cite{thorpe} and
Alexander~\cite{shlomo} are also based on the idea that a minimum
average coordination number is needed to prevent zero-frequency modes.
However, those theories are limited to non-rigid covalent networks or
systems interacting with attractive potentials, respectively.  By
contrast, the approach adopted here from Ref.~\cite{wyartthesis} can
be applied to low-coordination crystals, network glasses, as well as
repulsive glasses to explain why all these systems exhibit similar
structure in the density of vibrational states~\cite{wyartthesis}.

For our model glasses, we have shown that $(z_1 -z_c)/ z_c$ remains
small even when $z/z_c$ is arbitrarily large.  It is for this reason
that the marginal-coordination approach can be applied to these
systems with high coordination.  As long as $(z_1-z_c)/ z_c$ is small,
the behavior of the glass is governed by the physics of Point J and
isostaticity.  Even though the unjamming transition itself may be
inaccessible--as it is in the Lennard-Jones system studied here--the
effect of the longer-ranged part of the potential can be treated as a
correction.  This theory can therefore be viewed as a conceptual
generalization of the perturbation theory of liquids to the case of
jamming.

We thank Vincenzo Vitelli for stimulating discussions.  This work was
supported by DE-FG02-05ER46199 (AJL and NX) and DE-FG02-03ER46088 (SRN
and NX).


\begin{references}
\bibitem{A1dep} From p. 59 of Ref. [20], a better approximation may be
to use $k_{eff}$ instead of $k_1$, where $\langle (k_{eff}-k)/(2
k_{eff}+k)\rangle=0$, and the average is taken over all stiff
contacts.
\end{references}
\end{document}